\documentclass[twocolumn,superscriptaddress,preprintnumbers,amsmath,amssymb,prc,nofootinbib]{revtex4-2}
\usepackage[pdftex]{graphicx}
\usepackage{dcolumn}
\usepackage{bm}
\usepackage{ifpdf}
\usepackage{slashed}
\usepackage{amsfonts}
\usepackage{mathrsfs}
\usepackage{amssymb}
\usepackage{multirow}
\usepackage{CJK}
\usepackage[pdftex]{xcolor}
\def\lesssim{\ \raise.3ex\hbox{$<$}\kern-0.8em\lower.7ex\hbox{$\sim$}\ }
\def\gesim{\ \raise.3ex\hbox{$>$}\kern-0.8em\lower.7ex\hbox{$\sim$}\ }
\def\ket#1{\left|{#1}\right\rangle}
\def\braket#1#2{\left\langle{#1}\middle|{#2}\right\rangle}
\def\brakket#1#2#3{\left\langle{#1}\middle|{#2}\middle|{#3}\right\rangle}

\def\nuc#1#2#3{{}^{#2}_{#3}\mathrm{#1}}
\def\urm#1{\scriptstyle{\text{\textrm{\textmd{\textup{#1}}}}}}
\def\uurm#1{\scriptscriptstyle{\text{\textrm{\textmd{\textup{#1}}}}}}

\let\Re\relax
\let\Im\relax
\DeclareMathOperator{\Re}{Re}
\DeclareMathOperator{\Im}{Im}
\begin{document}
\begin{CJK*}{UTF8}{}
  \preprint{RIKEN-iTHEMS-Report-25}
  \title{Quartet correlations near the surface of $ N = Z $ nuclei}
  \author{Yixin Guo (\CJKfamily{gbsn}{郭一昕})}
  \email{guoyixin1997@ibs.re.kr}
  \affiliation{Center for Exotic Nuclear Studies, Institute for Basic Science, Daejeon 34126, Republic of Korea}
  \affiliation{
    Department of Physics, Graduate School of Science, The University of Tokyo,
    Tokyo 113-0033, Japan}
  \affiliation{
    RIKEN Center for Interdisciplinary Theoretical and Mathematical Sciences (iTHEMS),
    Wako 351-0198, Japan}
  \author{Tomoya Naito (\CJKfamily{min}{内藤智也})}
  \email{tnaito@ribf.riken.jp}
  \affiliation{
    RIKEN Center for Interdisciplinary Theoretical and Mathematical Sciences (iTHEMS),
    Wako 351-0198, Japan}
  \affiliation{
    Department of Physics, Graduate School of Science, The University of Tokyo,
    Tokyo 113-0033, Japan}
  \author{Hiroyuki Tajima (\CJKfamily{min}{田島裕之})}
  \email{hiroyuki.tajima@tnp.phys.s.u-tokyo.ac.jp}
  \affiliation{
    Department of Physics, Graduate School of Science, The University of Tokyo,
    Tokyo 113-0033, Japan}
     \affiliation{Quark Nuclear Science Institute, The University of Tokyo, Tokyo 113-0033, Japan}
  \author{Haozhao Liang (\CJKfamily{min}{梁豪兆})}
  \email{haozhao.liang@phys.s.u-tokyo.ac.jp}
  \affiliation{
    Department of Physics, Graduate School of Science, The University of Tokyo,
    Tokyo 113-0033, Japan}
  \affiliation{
    RIKEN Center for Interdisciplinary Theoretical and Mathematical Sciences (iTHEMS),
    Wako 351-0198, Japan}
    \affiliation{Quark Nuclear Science Institute, The University of Tokyo, Tokyo 113-0033, Japan}
  \date{\today}
  \begin{abstract}
    We theoretically investigate Cooper quartet correlations in $ N = Z $ doubly-magic nuclei
    ($ {}^{40} \mathrm{Ca} $, $ {}^{100} \mathrm{Sn} $, and $ {}^{164} \mathrm{Pb} $).
    We first examine the quartet condensation fraction in infinite symmetric nuclear matter by using the quartet Bardeen-Cooper-Schrieffer theory.
    Together with the total nucleon density profiles of doubly-magic nuclei obtained from the Skyrme Hartree-Fock calculation,
    we discuss the spatial distribution of quartet correlations in finite nuclei within the local density approximation.
    Large quartet condensate fractions are found at the surface region of an atomic nucleus due to the strong neutron-proton attractive interaction responsible for the deuteron formation in vacuum.
    Moreover, we discuss a possible microscopic origin of the Wigner term in the context of nucleon-quartet scattering in dilute symmetric nuclear matter.
    The nucleon-quartet scattering effect on the Wigner term is numerically estimated to be about one order of magnitude of the total empirical strength,
    indicating the importance of multinucleon clusters in the symmetry energy and mass formula in addition to the neutron-proton pairing.
  \end{abstract}
  \maketitle
\end{CJK*}

\section{Introduction}
\label{sec:I}
\par
Cluster formation has been one of the key issues in nuclear physics.
The Bardeen-Cooper-Schrieffer (BCS)-type pairing~\cite{
  Bardeen1957Phys.Rev.108.1175--1204}
of nucleons has greatly improved our understanding of nuclear structures and properties~\cite{
  Bohr1958PhysRev.110.936},
as the pair correlations in nuclei can be confirmed via odd-even staggering of nuclear masses~\cite{
  ring2004nuclear}.
\par
Recent experiments indicate the existence of light multinucleon clusters beyond pair formation.
The $ \alpha $-knockout reaction has revealed $ \alpha $-cluster formation in the surface region of medium-mass nuclei~\cite{
  Tanaka2021Science.371.260}.
Moreover, the Bose-Einstein condensate (BEC) of $ \alpha $ particles has been examined in excited $ \alpha $-conjugate nuclei~\cite{
  Tohsaki2001Phys.Rev.Lett.87.192501,
  adachi2021candidates,
  zhou20235,
  Wei2024Nuclsci,Ye2025NatRevPhys}.
Light cluster formation also plays a crucial role
in astrophysical environments such as a burning reactions inside stars, core-collapse supernovas,
and binary neutron-star mergers~\cite{
  Oertel2017RevModPhys.89.015007}.
As several light clusters, such as deuterons, tritons, and $ \alpha $ particles, are known to be formed in vacuum or dilute environments, it is important to develop many-body theories bridging dilute nuclear matter and nuclei to elucidate cluster formations and associated strong internucleon correlations in these systems. 
\par
In particular, an $ \alpha $ particle consisting of four nucleons is one of the most stable nuclear clusters, with the large binding energy $ E_{\alpha} = 28.29 \, \mathrm{MeV} $.
Accordingly, a statistical approach has been employed to describe in-medium $ \alpha $ particles in nuclei by including the Pauli-blocking effect~\cite{
  Typel2010Phys.Rev.C81.015803}.
However, in the context of strongly interacting fermions involving two-body cluster formations (i.e., BEC-BCS crossover)~\cite{
  Strinati2018Phys.Rep.738.1--76,
  Ohashi2020Prog.Part.Nucl.Phys.111.103739},
the interplay between the Pauli-blocking effect and the Cooper instability should be simultaneously taken into account for the cluster formation in many-body environments.
In this regard, many-body theories involving cluster formations beyond the BCS paradigm as well as the liquid-drop model have been examined in the context of four-body \textit{quartet} correlations, as in Refs.~\cite{
  Tohsaki2001Phys.Rev.Lett.87.192501,
  JSuper2010,
  PhysRevC.81.064310,
  PhysRevC.103.024316,
  PhysRevLett.131.193401}.  
So far, a variational ansatz based on the multiple occupation of quartet operators, called the quartet condensation model (QCM), has been proposed~\cite{
  FLOWERS1963586,
  Sandulescu2012Phys.Rev.C85.061303}.
Moreover, the extension of the BCS theory to quartet correlation,
called quartet BCS (QBCS) theory, has been developed and applied to nuclei~\cite{
  Baran2020Phys.Lett.B805.135462,
  Baran2020Phys.Rev.C102.061301}
and nuclear matter~\cite{
  Sen'kov2011Phys.Atom.Nuclei74.1267,
  Guo2022Phys.Rev.C105.024317}.
While the QBCS theory based on the quartet coherent state breaks particle-number conservation even in finite nuclei,
the results of the correlation energy agree well with those of QCM with particle-number conservation~\cite{Baran2020Phys.Lett.B805.135462}. 
In this sense, the QBCS theory can be a promising candidate for the description of Cooper quartet correlations even in finite nuclei. 
On the other hand, it is not so clear how the quartet correlations are spatially distributed in medium- or heavy-mass nuclei because of the computational difficulties.
\par
Meanwhile,
as the pairing correlations induce the odd-even staggering of nuclear binding energy, do the quartet correlations have impacts on nuclear properties in addition to the existence of the $ \alpha $-conjugate nuclei?
This reasonable question may be associated with a term whose microscopic origin is still unclear,
that is, the Wigner term in an empirical nuclear mass formula.
It is named after Wigner, who gave the first description in the investigation of spin-isospin symmetry of the nuclear force~\cite{
  Wigner1937Phys.Rev.51_106,
  Cassen1936Phys.Rev.50_846,
  Wigner1937Phys.Rev.51_947}.
The Wigner term is proportional to $ \left| N - Z \right| $ and is therefore distinct from the conventional symmetry energy term proportional to $ \left| N - Z \right|^2 $~\cite{
  Bethe1936Rev.Mod.Phys.8.82--229}.
Although nuclear binding energies can be well reproduced by mean-field approaches, the Wigner term still has to be added phenomenologically~\cite{
  Goriely2010Phys.Rev.C82.035804}
and its microscopical origin is under discussion. 
In finite nuclei, especially around the self-conjugate line ($ N = Z $), there are many related theoretical works to figure out the physical origin of the Wigner term via the investigation of proton-neutron pair correlations~\cite{
  Brenner1990Phys.Lett.B243.1--6,
  Zamfir1991Phys.Rev.C43.2879--2882,
  VanIsacker1995Phys.Rev.Lett.74.4607--4610,
  Myers1996Nucl.Phys.A601.141--167,
  Myers1997Nucl.Phys.A612.249--261,
  Satula1997Phys.Lett.B393.1--6,
  Satula1997Phys.Lett.B407.103--109,
  Jaenecke2002Phys.Rev.C66.024327,
  Jaenecke2005Phys.Lett.B605.87--94,
  Jaenecke2005Eur.Phys.J.A25.79--80,
  Bentley2013Phys.Rev.C88.014322},
and even some attempts for heavy nuclei~\cite{
  Zeldes1998Phys.Lett.B429.20--26,
  Zeldes1999J.Phys.G:Nucl.Part.Phys.25.949}.
Recently, a QCM study of the coexistence of quartets and pairs in even-even neutron-rich nuclei also showed a linear behavior of the correlation energy with respect to the extra neutron number~\cite{
  Sambataro2023Nucl.Phys.A1036.122675},
indicating that the quartet correlations may be related to the Wigner term.
Progress in producing heavy nuclei has also ignited interest in the Wigner term, and it gradually becomes directly observable up to larger mass number $ A $.
In this regard, it is also worth investigating whether or not the Wigner cusp discontinuity persists in heavier and even in yet unknown nuclei~\cite{
  Brenner1990Phys.Lett.B243.1--6,
  VanIsacker1995Phys.Rev.Lett.74.4607--4610}.
\par
To see the impact of quartet correlations on the nuclear equation of state, one may borrow knowledge of quantum mixtures by assuming that the present system is similar to nucleon-cluster mixtures. 
In condensed matter physics, quantum mixtures have been studied in terms of a polaron, which was originally proposed to describe electrons in ionic lattices~\cite{
  Landau1933Phys.Z.Sowjetunion3.664,
  landau1948effective}. 
Such a concept has been generalized to ultracold gas mixtures~\cite{
  baroni2024quantum},
where the polaron energy and equation of state can be measured experimentally~\cite{
  Schirotzek2009PhysRevLett.102.230402,
  navon2010equation}.
Recently, the notion of polarons has also been extended to nuclear problems involving cluster formations~\cite{
  Nakano2020Phys.Rev.C102.055802,
  Moriya2021Phys.Rev.C104.065801,
  tajima2024polaronic,
  tajima2024intersections,
  tajima2024polaronicneutrondilutealpha}.
Indeed, the polaron energy of protons, proportional to proton number, in asymmetric nuclear matter is found to have a close relation to the nuclear symmetry energy~\cite{
  tajima2024intersections}.
This indicates that the linear increase of the energy with respect to the particle number in quantum mixtures is connected with the intercomponent interaction.
Accordingly, if one can find the spatial profile of quartet clusters and the interaction between such clusters and nucleons, one can estimate a similar contribution associated with the Wigner term in symmetry energy as a consequence of quartet correlations.
\par
In this paper, we theoretically investigate how quartet correlations emerge in $ N = Z $ doubly-magic nuclei in the following way:
We first obtain the quartet condensation fraction for infinite symmetric nuclear matter by the QBCS theory.
Together with the total nucleon density profiles of doubly-magic nuclei from the Skyrme Hartree-Fock calculation,
we perform the extension to the corresponding finite nuclei with the local density approximation (LDA).
We show the spatial distributions of the quartet condensation fraction in several nuclei, that is, $ \nuc{Ca}{40}{} $, $ \nuc{Sn}{100}{} $, and $ \nuc{Pb}{164}{} $.
Here, we choose these $ N = Z $ doubly-magic nuclei to be treated as consisting of quartets for the following investigation of nucleon-quartet correlation.
The quartet condensation fraction is found to be localized near the surface of nuclei.
With such a result, we examine the role of nucleon-cluster scattering in the energy density
by considering the $ s $-wave repulsion between a nucleon and an $ \alpha $ particle relevant to the low-density regime,
i.e., the surface region of nuclei.
We show that the contribution to the Wigner term can arise from the nucleon-quartet correlation effect and we estimate its impact in the aforementioned nuclei,
where a doubly-magic nucleus is treated as a core consisting of quartets, and valence neutrons are considered.
\par
This paper is organized as follows.
In Sec.~\ref{sec:II}, we introduce our formalism including the QBCS theory combined with Skyrme Hartree-Fock calculation and LDA.
Moreover, the effects of nucleon-cluster scattering and the relation to the Wigner term are presented.
In Sec.~\ref{sec:IV}, we present the numerical results of the quartet condensation fraction for infinite symmetric nuclear matter via the QBCS theory and the spatial profile of the quartet condensation fraction in nuclei by using the Skyrme Hartree-Fock calculation with LDA.
Furthermore, we estimate the Wigner-type energy contribution originating from the coexistence of quartets and nucleons in the surface region. 
Finally, a summary and perspectives are given in Sec.~\ref{sec:V}.
In the following, we take $ \hbar = c = k_{\urm{B}} = 1 $.
\section{Formalism}
\label{sec:II}
\par
In this section, we present the formalism of the QBCS theory and its extension to finite nuclei with LDA.
Upon such a framework, we consider the nucleon-quartet correlations, which are not included in the QBCS theory.
\subsection{QBCS theory}
\label{sec:IIA}
\par
First, we review the QBCS theory in infinite symmetric nuclear matter.
For convenience, the system size is taken to be a unit for the infinite-matter calculation.
For detailed derivations, see Refs.~\cite{
  Guo2022Phys.Rev.C105.024317,
  Guo2022Phys.Rev.Research4.023152}.
\par
Since we focus on the quartet correlations near the surface of finite nuclei,
we consider only the isoscalar pairing in this work, which is important in the low-density regime. 
The corresponding Hamiltonian is given by
\begin{align}
  \label{hamiltonian}
  H
  & = 
    \sum_{\bm{p}, s_z}
    \left(
    \varepsilon_{\nu, \bm{p}}
    \nu_{\bm{p}, s_z}^{\dagger}
    \nu_{\bm{p}, s_z}
    +
    \varepsilon_{\pi, \bm{p}}
    \pi_{\bm{p}, s_z}^{\dagger}
    \pi_{\bm{p}, s_z}
    \right)
    \notag \\
  & \quad
    +
    \frac{1}{2}
    \sum_{\bm{P}, \bm{q}, \bm{q}'}
    \sum_{S_z = -1}^{+1}
    \mathcal{D}_{1, S_z}^{\dagger} \left( \bm{P}, \bm{q}  \right)
    \mathcal{V}                    \left( \bm{q}, \bm{q}' \right)
    \mathcal{D}_{1, S_z}           \left( \bm{P}, \bm{q}' \right).
\end{align}
The parameters introduced in the Hamiltonian are summarized in Table~\ref{tab:para}.
In this paper, we take $ M_{\nu} = M_{\pi} \equiv M $.
\begin{table*}[tb]
  \centering
  \caption{
    Summary of the parameters in the Hamiltonian given by Eq.~\eqref{hamiltonian}.}
  \label{tab:para}
  \begin{ruledtabular}
    \begin{tabular}{ll}
      Parameter                                                   & Physical meaning \\
      \hline 
      $ \nu^\dagger $                                             & Neutron creation operator \\
      $ \pi^\dagger $                                             & Proton creation operator \\
      $ \bm{p} $                                                  & Single-particle momentum \\
      $ \bm{q} = \left( \bm{p}_1 - \bm{p}_2 \right)/2 $ & Relative momentum \\
      $ \bm{P} = \bm{p}_1 + \bm{p}_2 $                            & Center-of-mass momentum \\
      $ S_z $                                                     & Third component of spin (or $ s_z $ for single nucleon) \\
      $ t_3 $                                                     & Third component of isospin for single nucleon ($ +1/2 $ for neutrons and $ -1/2 $ for protons) \\
      $ \mathcal{V} $                                             & Interaction strength in the isoscalar channel \\
      $ \varepsilon_{i, \bm{p}} = {\bm{p}^{2}}/{2 M_i}-\mu_i $ ($ i = \pi $, $ \nu $)
                                                                  & Single-particle energy \\
      $ \mu_i $                                                   & Nucleon chemical potential \\
      $ M_i $                                                     & Nucleon mass \\
    \end{tabular}
  \end{ruledtabular}
\end{table*}
\par
The two-nucleon annihilation operators are defined by
\begin{align}
  \mathcal{D}_{1, S_z}(\bm{P}, \bm{q})
  & =
    \sum_{s_z, s'_z}
    \sum_{t_3, t'_3}
    \braket{\frac{1}{2} \frac{1}{2} s_z s'_z}{1 S_z}
    \braket{\frac{1}{2} \frac{1}{2} t_3 t'_3}{00}
    \notag \\
  & \quad 
    \times
    c_{\frac{\bm{P}}{2} -\bm{q}, s_z,  t_3}
    c_{\frac{\bm{P}}{2} +\bm{q}, s'_z, t'_3},
\end{align}
where
$ \braket{\frac{1}{2} \frac{1}{2} s_z s'_z}{1 S_z} $ and 
$ \braket{\frac{1}{2} \frac{1}{2} t_3 t'_3}{00} $
are the Clebsch-Gordan coefficients in the spin and isospin spaces, respectively.
The corresponding creation operators are their conjugates.
Moreover, we assume the symmetry of the interaction as
$ \mathcal{V} \left(  \bm{q},   \bm{q}' \right)
=
\mathcal{V} \left(  \bm{q}, - \bm{q}' \right)
=
\mathcal{V} \left( -\bm{q},   \bm{q}' \right)
=
\mathcal{V} \left( -\bm{q},  -\bm{q}' \right) $.
\par
Considering the zero center-of-mass momentum $ \bm{P} = \bm{0} $ and taking the case of $ \bm{q} = \bm{q}' $,
the quartet creation operator can be introduced as
\begin{equation}
  \label{eq:alpha_creation}
  \alpha^\dagger_{\bm{q}}
  =
  \sum_{S_z, S'_z}
  \braket{11 S_z S'_z}{00}
  \mathcal{D}_{1, S_z }^\dagger \left( 0, \bm{q} \right)
  \mathcal{D}_{1, S'_z}^\dagger \left( 0, \bm{q} \right).
\end{equation}
\begin{widetext}
  The QBCS trial wave function is taken as
  \begin{equation}
    \label{twf1}
    \ket{\Psi_{\urm{QBCS}}}
    =
    \prod_{\bm{q}}
    \left[
      u_{\bm{q}}
      +
      \frac{\sqrt{2}}{2}
      \sum_{S_z}
      v_{\bm{q}, S_z}
      \mathcal{D}_{1, S_z}^{\dagger} \left( 0, \bm{q} \right)
      +
      \frac{1}{2}
      w_{\bm{q}}
      \alpha^{\dagger}_{\bm{q}}
    \right]
    \ket{0}
  \end{equation}
  with the normalization
  \begin{equation}
    \left| u_{\bm{q}} \right|^2
    +
    \left| \bm{v}_{\bm{q}} \right|^2
    +
    \left| w_{\bm{q}} \right|^2
    =
    1,
  \end{equation}
  where we introduced
  $ \left| \bm{v}_{\bm{q}} \right|^2 = \sum_{S_z} \left| v_{\bm{q}, S_z} \right|^2 $.
  Calculating the expectation value of Hamiltonian
  $ \brakket{\Psi_{\urm{QBCS}}}{H}{\Psi_{\urm{QBCS}}} $
  and the pairing energy gaps
  \begin{equation}
    \label{gapeq}
    \Delta_{\bm{q}, S_z}^{\urm{QBCS}}
    =
    -
    \sum_{\bm{q}'}
    \mathcal{V} \left( \bm{q}, \bm{q}' \right)
    \left[
      u_{\bm{q}'}^*
      v_{\bm{q}', S_z}
      +
      \delta_{S_z, +1}
      v_{\bm{q}', - S_z}^*
      w_{\bm{q}'}
      +
      \delta_{S_z, -1}
      v_{\bm{q}', - S_z}^*
      w_{\bm{q}'}
      -
      \frac{1}{2}
      \delta_{S_z, 0}
      \left(
        v_{\bm{q}', -S_z}^*
        w_{\bm{q}'}
        +
        v_{\bm{q}', -S_z}^*
        w_{- \bm{q}'}
      \right)
    \right],   
  \end{equation}
  we apply the variational principle leading to
  \begin{subequations}
    \begin{align}
      v_{\bm{q}, +1}
      & = 
        \frac{u_{\bm{q}} \Delta_{\bm{q}, +1}^{\urm{QBCS}} + w_{\bm{q}} {\Delta^*}_{\bm{q}, -1}^{\urm{QBCS}}}{B_{\bm{q}} + \left( \varepsilon_{\pi, \bm{q}} + \varepsilon_{\nu, -\bm{q}} \right)},
        \quad
        v_{\bm{q}, -1}
        =
        \frac{u_{\bm{q}} \Delta_{\bm{q}, -1}^{\urm{QBCS}} + w_{\bm{q}} {\Delta^*}_{\bm{q}, +1}^{\urm{QBCS}}}{B_{\bm{q}} + \left( \varepsilon_{\pi, \bm{q}} + \varepsilon_{\nu, -\bm{q}} \right)},
        \label{v1} \\
      v_{\bm{q}, 0}
      & = 
        \frac{u_{\bm{q}} \Delta_{\bm{q}, 0}^{\urm{QBCS}} - \frac{1}{2} \left( w_{\bm{q}} + w_{-\bm{q}} \right) {\Delta^*}_{\bm{q}, 0}^{\urm{QBCS}}}{B_{\bm{q}} + \left( \varepsilon_{0, \bm{q}} + \varepsilon_{0, -\bm{q}} \right)},
        \label{v0} \\
      w_{\bm{q}}
      & =
        \frac{v_{\bm{q}, -1} \Delta_{\bm{q}, +1}^{\urm{QBCS}} + v_{\bm{q}, +1} \Delta_{\bm{q}, -1}^{\urm{QBCS}} - \frac{1}{2} \left( v_{\bm{q}, 0} + v_{-\bm{q}, 0} \right) \Delta_{\bm{q}, 0}^{\urm{QBCS}}}{B_{\bm{q}} + 2 \left( \varepsilon_{\pi, \bm{q}} + \varepsilon_{\nu, -\bm{q}} \right)}
        \label{eq:w_q}
    \end{align}
  \end{subequations}
  with $ \varepsilon_{0, \bm{q}} = \left( \varepsilon_{\nu, \bm{q}} + \varepsilon_{\pi,\bm{q}} \right) / 2 $
  and
  \begin{equation}
    B_{\bm{q}}
    =
    \frac{1}{2u_{\bm{q}}}
    \sum_{S_z}
    \left(
      v_{\bm{q}, S_z}^*
      \Delta_{\bm{q}, S_z}^{\urm{QBCS}}
      +
      v_{\bm{q}, S_z}
      {\Delta^*}_{\bm{q}, S_z}^{\urm{QBCS}}
    \right)
    =
    \frac{1}{u_{\bm{q}}}
    \Re
    \left(
      \bm{v}_{\bm{q}}^*
      \cdot
      \bm{\Delta}_{\bm{q}}^{\urm{QBCS}}
    \right).
  \end{equation}
\end{widetext}
\par
Similarly to the case of the conventional BCS theory~\cite{
  Salasnich2005Phys.Rev.A72.023621}, 
the quartet condensate density and its fraction with respect to the total density
can be calculated as
\begin{align}
  \rho_{\alpha}
  & =
    \sum_{\bm{q}}
    \left|
    \brakket{\Psi}{\alpha \left( \bm{q} \right)}{\Psi}
    \right|^2, \\
  \mathcal{C}
  & =
    \frac{4 \rho_{\alpha}}{\rho}
    =
    \frac{4 \sum_{\bm{q}} u_{\bm{q}}^2 w_{\bm{q}}^2}{\sum_{\bm{q}} \left( 2 \left| v_{\bm{q}} \right|^2 + 4 \left| w_{\bm{q}} \right|^2 \right)}.
\end{align}
Instead of the contact-type nucleon-nucleon interaction adopted in the previous work~\cite{
  Guo2022Phys.Rev.C105.024317},
here we employ the finite-range interaction with the Gaussian form factor given by
\begin{equation}
  \mathcal{V} \left( \bm{q}, \bm{q}' \right)
  =
  -
  \lambda
  e^{- \left( q^2 + q'^2 \right) /2b}
\end{equation}
with the relative momenta $ \bm{q} $ ($ \bm{q}' $) of paired nucleons,
the coupling constant $ \lambda $, and the range parameter $ b $~\cite{
  Roepke1998Phys.Rev.Lett.80.3177--3180}.
These parameters 
are taken as $ \lambda = 807.5 \, \mathrm{MeV} \cdot \mathrm{fm}^3 $ and $ b = 1.50 \, \mathrm{fm}^{-1} $,
respectively, to reproduce the $ \alpha $ particle binding energy $ E_{\alpha} = 28.29 \, \mathrm{MeV} $ in free space.
\par
  Before concluding the QBCS theory, we would like to elaborate more on the difference between Cooper quartets and bound $\alpha$ particles.
  In the nuclear systems, with increasing density, the bound $\alpha$ states are gradually modified by the in-medium effect and eventually disappear at a critical density.
  Such a critical point is referred as the Mott density of $\alpha$ particles~\cite{Roepke1982Nucl.Phys.A379.536--552,Roepke1983Nucl.Phys.A399.587--602,Roepke1998Phys.Rev.Lett.80.3177--3180,Beyer2000Phys.Lett.B488.247--253}.
  The Cooper quartets investigated in the present work may be identical to the bound $\alpha$ particles only in the dilute regime where the density is close to zero.
  If the density becomes close to or even above the Mott density, the Cooper quartet should become distinct from the bound $\alpha$, where an $\alpha$ particle is absent due to the medium breakup effect.
  In other words, the Cooper quartet would be an in-medium BCS-like four-body correlation which exists even above the Mott density.
  In this paper, we call such four-body bound states and correlations as quartets in the entire density regime. 
  Moreover, it should be noted that we perform the investigation in a similar manner as the conventional BCS theory, which takes condensed pairs into account while disregarding uncondensed pairs.
  Based on that, the quartet condensate density is not equal to the total quartet density because of possible uncondensed quartets.
  Note that, it is known that uncondensed fractions are relatively small in the pairing case~\cite{Ohashi2020Prog.Part.Nucl.Phys.111.103739} in the weak-coupling regime (corresponding to higher densities).
\subsection{LDA extension to finite nuclei}
\par
Once we have the total nucleon density profiles of finite nuclei,
we can apply the LDA to calculate the corresponding spatial distribution of the quartet condensation fraction.
In detail, for a certain nucleus, at a fixed spatial coordinate $ r $,
the local quartet condensation fraction $ \mathcal{C} \left( r \right) $ is approximated by the one obtained
in infinite symmetric nuclear matter with the same total local nucleon density $ \rho \left( r \right) $ via the QBCS theory as introduced in Sec.~\ref{sec:IIA}.
\par
To this end, we perform the Skyrme Hartree-Fock calculation~\cite{
  PhysRevC.5.626} with the SLy4 energy density functional~\cite{
  CHABANAT1997710}.
The spherical symmetry is assumed and the spatial coordinate is discretized.
A $ 160 \times 0.1 \, \mathrm{fm} $ box is used for the radial coordinate to perform the numerical calculation.
Throughout this calculation, we obtain the density profile of an atomic nucleus.
\par
After obtaining the quartet condensation fractions of finite nuclei as functions of spatial coordinates,
we will discuss the quartet correlations at the surface of $ N = Z $ nuclei.
\subsection{Neutron self-energy for the $ s $-wave neutron-cluster interaction}
\par
In order to discuss the nucleon-quartet correlations in the dilute regime,
which are not taken into account in the QBCS theory, we consider a system consisting of quartets and a few residual neutrons.
For simplicity, we assume that the effective nucleon-quartet interaction can be expressed in terms of the nucleon-$ \alpha $ interaction.
In such a case, the $ s $-wave neutron-$ \alpha $ interaction is characterized by the finite-range repulsive interaction
\begin{align}
  \label{eq:v_na}
  V_{n \alpha}
  & = 
    \sum_{s_z}
    \sum_{\bm{k}, \bm{k}', \bm{P}}
    U \left( \bm{k}, \bm{k}' \right)
    \notag \\
  & \quad
    \times
    \nu_{\bm{k}  + \bm{P} / 5, s_z}^{\dagger}
    \alpha_{- \bm{k}  + 4 \bm{P} / 5}^{\dagger}
    \alpha_{- \bm{k}' + 4 \bm{P} / 5}
    \nu_{\bm{k}' + \bm{P} / 5, s_z},
\end{align}
where $ \alpha_{\bm{k}}^{\text{($ \dagger $)}} $ is an annihilation (creation) operator of an $ \alpha $-like quartet cluster as in the QBCS theory.
We employ the separable form of $ U \left( \bm{k}, \bm{k}' \right) $ as
\begin{equation}
  U \left( \bm{k}, \bm{k}' \right)
  =
  U_0
  \gamma_k
  \gamma_{k'},
\end{equation}
with the form factor~\cite{Mongan1969PhysRev.178.1597}
\begin{equation}
  \gamma_{k}
  =
  \frac{\sqrt{1 + \chi k^2}}{1 + \left( k / \Lambda \right)^2}.
\end{equation}
The parameters $ U_0 $, $ \chi $, and $ \Lambda $ are determined in such a way that the scattering length
$ a = 2.64 \, \mathrm{fm} $ and the effective range $ r = 1.43 \, \mathrm{fm} $~\cite{
  Kanada1979Prog.Theor.Phys.61.1327--1341,
  Nakano2020Phys.Rev.C102.055802}
are reproduced in the low-energy scattering (for more details, see Appendix~\ref{app:a}).
\par
We then obtain the interaction effects on a nucleon coexisting with the quartet condensate based on the Green's function formalism.
The nucleon Green's function reads
\begin{equation}
  \label{eq:17}
  G \left( \bm{k}, \omega \right)
  =
  \frac{1}{\omega - k^2 / \left( 2M \right) - \Sigma \left( \bm{k}, \omega \right) + i0}.
\end{equation}
Here, we consider the Beliaev-type contribution~\cite{
  Rath2013PhysRevA.88.053632}
for the self-energy as
\begin{equation}
  \Sigma \left( \bm{k}, \omega \right)
  =
  T \left( 4\bm{k}/5, 4\bm{k}/5; \omega \right)
  \rho_{\alpha},
\end{equation}
where
\begin{equation}
  \label{eq:tmat}
  T \left( \bm{k}, \bm{k}'; \omega \right)
  =
  U_0
  \gamma_k
  \gamma_{k'}
  \left[
    1
    -
    U_0
    \sum_{\bm{p}}
    \frac{\gamma_p^2}{\omega + i0 - p^2/2M_{\urm{r}}}
  \right]^{-1}
\end{equation}
is the neutron-quartet scattering $ T $-matrix with the reduced mass $ M_{\urm{r}} = 4M/5 $ (see also Appendix~\ref{app:a}),
and
$ \rho_{\alpha} $ is the quartet condensate density.
While we do not consider the contribution from noncondensed quartet clusters,
such excited states can be regarded as in-medium nucleons resulting from the breakup of clusters due to the Pauli-blocking effect. 
Also, since we are interested in the low-density regime,
the Pauli-blocking of nucleons in the intermediate state is neglected in Eq.~\eqref{eq:tmat}.
\par
The self-energy shift of a dressed neutron is obtained as
\begin{equation}
  \label{eq:ep}
  E_{\urm{P}}
  =
  \Re
  \Sigma \left( \bm{0}, E_{\urm{P}} \right)
  =
  \Re
  T \left( \bm{0}, \bm{0}; E_{\urm{P}} \right)
  \rho_{\alpha}.
\end{equation}
At the weak-coupling limit, we find
\begin{equation}
  \label{eq:e_h}
  E_{\urm{P}}
  \simeq
  T \left( \bm{0}, \bm{0}; 0 \right)
  \rho_{\alpha}
  =
  \frac{2 \pi a}{M_{\urm{r}}}
  \rho_{\alpha}
  \equiv
  E_{\urm{H}},
\end{equation}
where the Hartree shift $ E_{\urm{H}} $ has been reproduced exactly. 
We note that the Fock term is absent here because $ V_{n \alpha} $ works between two distinguishable objects (i.e., nucleon and $ \alpha $-like quartet cluster).
\par
For low-energy excitations,
Eq.~\eqref{eq:17} can approximately be simplified as
\begin{equation}
  G \left( \bm{k}, \omega \right)
  \simeq
  \frac{Z_{\urm{P}}}{\omega - k^2/ \left( 2 M_{\urm{eff}} \right) - E_{\urm{P}} - i \Gamma_{\urm{P}}/2},
\end{equation}
where the quasiparticle residue
\begin{equation}
  Z_{\urm{P}}
  =
  \left[
    1
    -
    \Re
    \left.
      \frac{\partial \Sigma \left( \bm{k}, \omega \right)}{\partial \omega}
    \right|_{\omega=E_{\uurm{P}}}
  \right]^{-1},
\end{equation}
the inverse effective mass
\begin{equation}
  \frac{M}{M_{\urm{eff}}}
  =
  Z_{\urm{P}}
  \left[
    1
    +
    M
    \Re
    \left.
      \frac{\partial^2 \Sigma \left( \bm{0}, E_{\urm{P}} \right)}{\partial k^2}
    \right|_{k = 0}
  \right],
\end{equation}
and the decay width
\begin{equation}
  \Gamma_{\urm{P}}
  =
  -
  2Z_{\urm{P}}
  \Im
  \Sigma \left( \bm{0}, E_{\urm{P}} \right).
\end{equation}
\begin{figure}[t]
  \centering
  \includegraphics[width=1.0\linewidth]{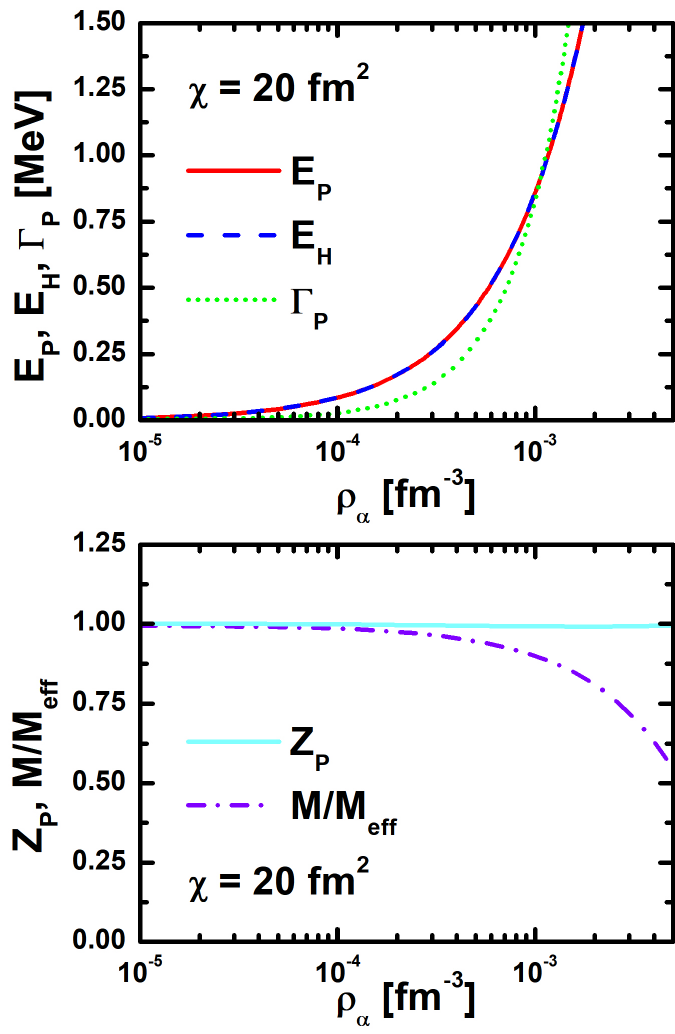}
  \caption{
    Quasiparticle parameters of a neutron in the quartet condensate.
    The interaction parameter is taken as $ \chi = 20 \, \mathrm{fm}^2 $. 
    Here, $ E_{\urm{P}} $, $ E_{\urm{H}} $, and $ \Gamma_{\urm{P}} $ are the Baeliaev-type self-energy, Hartree shift, and decay width, respectively.
    In the lower panel, the quasiparticle residue $ Z_{\urm{P}} $ and the inverse effective mass $ M / M_{\urm{eff}} $ are plotted.}
  \label{fig:chi20}
\end{figure}
\par
In Fig.~\ref{fig:chi20}, we show $ E_{\urm{P}} $, $ Z_{\urm{P}} $, $ M/ M_{\urm{eff}} $, and $ \Gamma_{\urm{P}} $
as functions of the quartet condensation density $ \rho_{\alpha} $,
where we take $ \chi = 20 \, \mathrm{fm}^2 $.
For comparison, $ E_{\urm{H}} $ given by Eq.~\eqref{eq:e_h} is also plotted. 
One can find that $ E_{\urm{P}} $ given by Eq.~\eqref{eq:ep} agrees well with $ E_{\urm{H}} $ in the low-density region shown in Fig.~\ref{fig:chi20}. 
In addition, we obtain $ \Gamma_{\urm{P}} < E_{\urm{P}} $ and $ M_{\urm{eff}} \simeq M $ at
$ \rho_{\alpha} \lesssim 10^{-3} \, \mathrm{fm} $,
indicating that $ E_{\urm{H}} $ without further Fermi-liquid corrections gives a qualitatively good estimation of the correlation energy for nucleon-cluster scattering.
We note that the present effective mass $ M_{\urm{eff}} $ is defined by expanding the self-energy at zero momentum, and it is different from the conventional studies~\cite{LI201829} where the self-energy is expanded near the nucleon Fermi momentum.
Our result in Fig.~\ref{fig:chi20} shows that $ M_{\urm{eff}} $ is larger than $ M $ in the dilute regime due to the nucleon-$ \alpha $ interaction, consistent with a cold-atomic counterpart~\cite{
  Rath2013PhysRevA.88.053632}.
Such a tendency can also be found in proton excitations in dilute asymmetric nuclear matter with the strong proton-neutron interaction~\cite{
  BAYM1971225,tajima2024polaronic}.
Incidentally, the $ p $-wave resonance associated with $ \nuc{He}{5}{} $ resonant states,
which plays an important role in the formation of halo nuclei~\cite{
  Hammer_2017},
additionally enhances $ M_{\urm{eff}} $ at larger $ \rho_{\alpha} $ but
does not contribute to $ E_{\urm{P}} $~\cite{
  tajima2024polaronicneutrondilutealpha}.
\par
Meanwhile,
the Landau-Pomeranchuk-type energy density~\cite{
  PhysRevLett.100.030401}
of residual neutrons with the quartet condensates reads
\begin{equation}\label{eq24}
  \mathcal{E} \left( \rho \right)
  =
  \mathcal{E}_{\alpha} \left( 4 \rho_{\alpha} \right)
  +
  E_{\urm{P}}
  \,
  {\Delta \rho_n}
  +
  \frac{\left( 3 \pi^2 \right)^{5/3}}{10 \pi^2 M_{\urm{eff}}}
  {\Delta \rho_n}^{5/3}
  +
  \cdots
\end{equation}
where $ \mathcal{E}_{\alpha} \left( 4 \rho_{\alpha} \right) $ is the ground-state energy density of the quartet condensate.
Here, $ {\Delta \rho_n} $ is the residual neutron density,
and we denote the total nucleon density as $ \rho = 4 \rho_{\alpha} + {\Delta \rho_n} $,
and approximate the energy density of symmetric nuclear matter by
\begin{equation}
  \mathcal{E}_{\urm{SNM}} \left( \rho \right)
  \simeq
  \mathcal{E}_{\alpha} \left( 4 \rho_{\alpha} + {\Delta \rho_n} \right).
\end{equation}
Expanding the right-hand side with respect to $ {\Delta \rho_n} $,
we find
\begin{equation}
  \mathcal{E}_{\alpha} \left( 4 \rho_{\alpha} + {\Delta \rho_n} \right)
  =
  \mathcal{E}_{\alpha} \left( 4 \rho_{\alpha} \right)
  +
  \frac{1}{4}
  \mu_{\alpha}
  \,
  {\Delta \rho_n}
  +
  O \left( {\Delta \rho_n}^2 \right),
\end{equation}
where
$ \mu_{\alpha} = \frac{\partial \mathcal{E}_{\alpha}}{\partial \rho_{\alpha}} $ 
is the effective chemical potential of an $ \alpha $-like quartet cluster.
Eventually, we find
\begin{equation}
  \label{eq27}
  \mathcal{E} \left( \rho \right)
  \simeq 
  \mathcal{E}_{\alpha} \left( 4 \rho_{\alpha} + {\Delta \rho_n} \right)
  -
  \frac{1}{4}
  \mu_{\alpha}
  \,
  {\Delta \rho_n}
  +
  E_{\urm{P}}
  \,
  {\Delta \rho_n}.
\end{equation}
The symmetry energy is defined by~\cite{
  tajima2024intersections}
\begin{align}
  \mathcal{S}
  & \equiv
    \frac{\mathcal{E} \left( \rho \right)}{\rho}
    -
    \frac{\mathcal{E}_{\urm{SNM}} \left( \rho \right)}{\rho}
    \notag \\
  & =
    \left(
    E_{\urm{P}}
    -
    \frac{\mu_{\alpha}}{4}
    \right)
    \frac{{\Delta \rho_n}}{\rho}
    +
    O \left( {\Delta \rho_n}^{5/3} \right).
    \label{eq:epol}
\end{align}
As a first approximation, one may assume $ \mu_{\alpha} \simeq 0 $
(i.e., ignoring the quartet-quartet interaction) and obtain
$ \mathcal{S} \simeq E_{\urm{P}} \, {\Delta \rho_n} / \rho $.
While we consider residual neutrons explicitly,
the same equation can be obtained for the residual proton case by assuming that the effects of the interaction between residual nucleons are included in the QBCS theory.
\par
On the other hand, the linear behavior of $ \mathcal{S} $ with respect to the residual nucleon density is reminiscent of
the Wigner term.
As for the neutron-rich matter with a small asymmetry, the corresponding equation of state expanded up to the leading order can be generally given by
\begin{align}
  \label{eq:enucl}
  \frac{\mathcal{E} \left( \rho \right)}{\rho}
  & = 
    \frac{\mathcal{E}_{\urm{SNM}} \left( \rho \right)}{\rho}
    +
    W
    \frac{\rho_{\urm{$ n $, total}} - \rho_{\urm{$ p $, total}}}{\rho}
    +
    \cdots
    \notag \\
  & =
    \frac{\mathcal{E}_{\urm{SNM}} \left( \rho \right)}{\rho}
    +
    W
    \frac{{\Delta \rho_n}}{\rho}
    +
    \cdots,
\end{align}
where $ \rho_{\urm{$ n $ ($ p $), total}} $ is the total neutron (proton) density.
Comparing Eqs.~\eqref{eq:epol} and \eqref{eq:enucl},
one can obtain $ W \simeq E_{\urm{P}} \simeq E_{\urm{H}} $ for small residual nucleon densities,
where we assumed that the density asymmetry
$ \rho_{\urm{$ n $, total}} - \rho_{\urm{$ p $, total}} $
is directly related to the residual nucleon density $ {\Delta \rho_n} $
(otherwise, the neutron-proton pairing can occur but such an effect has already been taken into account in the QBCS theory).
\section{Numerical results}
\label{sec:IV}
\begin{figure}[t]
  \centering
  \includegraphics[width=1.0\linewidth]{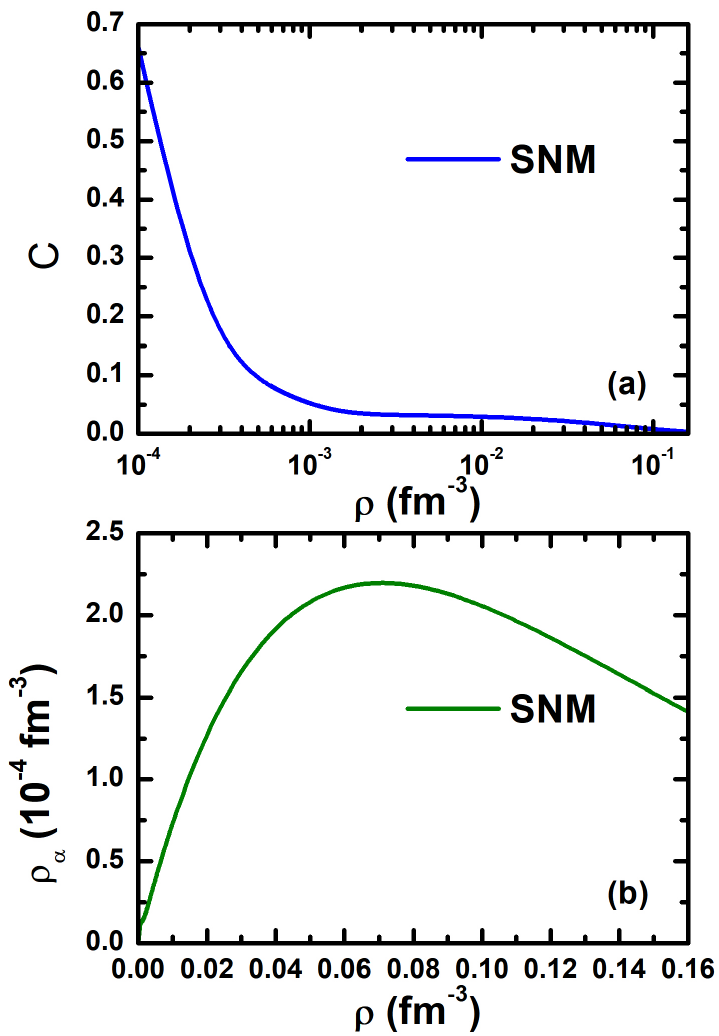}
  \caption{
    (a)~Quartet condensation fraction $ \mathcal{C} $ in infinite symmetric nuclear matter
    as a function of total density obtained from the QBCS theory~\cite{
      Guo2022Phys.Rev.C105.024317}.
    (b)~Same as upper panel but for $ \rho_{\alpha} = \mathcal{C} \rho / 4 $.}
  \label{fig:qcf}
\end{figure}
\par
First, we present the numerical results of the quartet condensate fraction $ \mathcal{C} $ obtained by the QBCS theory in infinite symmetric nuclear matter.
Figure~\ref{fig:qcf}(a) shows $ \mathcal{C} $ as a function of the total nucleon density $ \rho $. 
In the dilute limit, $ \mathcal{C} $ approaches $1$ (although we do not show it explicitly)
and immediately decrease up to $ \rho \simeq 10^{-3} \, \mathrm{fm}^{-3} $.
This result reflects that, in the low-density regime,
the formation of $ \alpha $ particles is dominant and forms the condensate because we use the zero-temperature formalism.
In this regard, as shown in Fig.~\ref{fig:qcf}(b), the quartet condensate density $ \rho_{\alpha} $ given by $ \mathcal{C} \rho / 4 $ itself is not so large.
\par
At larger densities, $ \mathcal{C} $ decreases relatively slowly with increasing $ \rho $.
This indicates that the interaction energy is gradually smaller than the nucleon Fermi energy and the system is in the weak-coupling regime as in the case of the density-induced BEC-BCS crossover~\cite{
  Tajima2022Phys.Rev.A106.043308}.
On the other hand, the quartet condensate density given by
$ \rho_{\alpha} = \mathcal{C} \rho / 4 $ [see Fig.~\ref{fig:qcf}(b)] increases up to
$ \rho \simeq 0.07 \, \mathrm{fm}^{-3} $, and starts to decrease with increasing $ \rho $.
This behavior is associated with the finite-range properties of the nucleon-nucleon interaction.
As in the pairing case~\cite{
  Jin2010Phys.Rev.C82.024911,
  Tajima2019Sci.Rep.9.18477},
the effective-range correction reduces the cluster formation at high densities.
Eventually, the maximum quartet condensate density can be estimated as
$ \rho_{\alpha} \simeq 2.2 \times 10^{-4} \, \mathrm{fm}^{-3} $,
implying that the low-energy assumption about the nucleon-cluster correlations characterized by Eq.~\eqref{eq:v_na} could be valid.
\par
We note that while Fig.~\ref{fig:qcf}(a) shows a qualitatively reasonable description of the quartet condensation fraction, there are still some improvements left for future work.
In particular, the relatively slow decrease of $ \mathcal{C} $ in the high-density regime may be related to the overestimation of $ \mathcal{C} $ in the present QBCS theory without the short-range repulsion and the three-body forces.
Nevertheless, 
the quartet condensation fraction generally exhibits a small value in the high-density regime due to the finite-range property of $ \mathcal{V} \left( \bm{q}, \bm{q}' \right) $, and hence our results would be unchanged qualitatively by this overestimation.
\begin{figure*}
  \centering
  \includegraphics[width=1.0\linewidth]{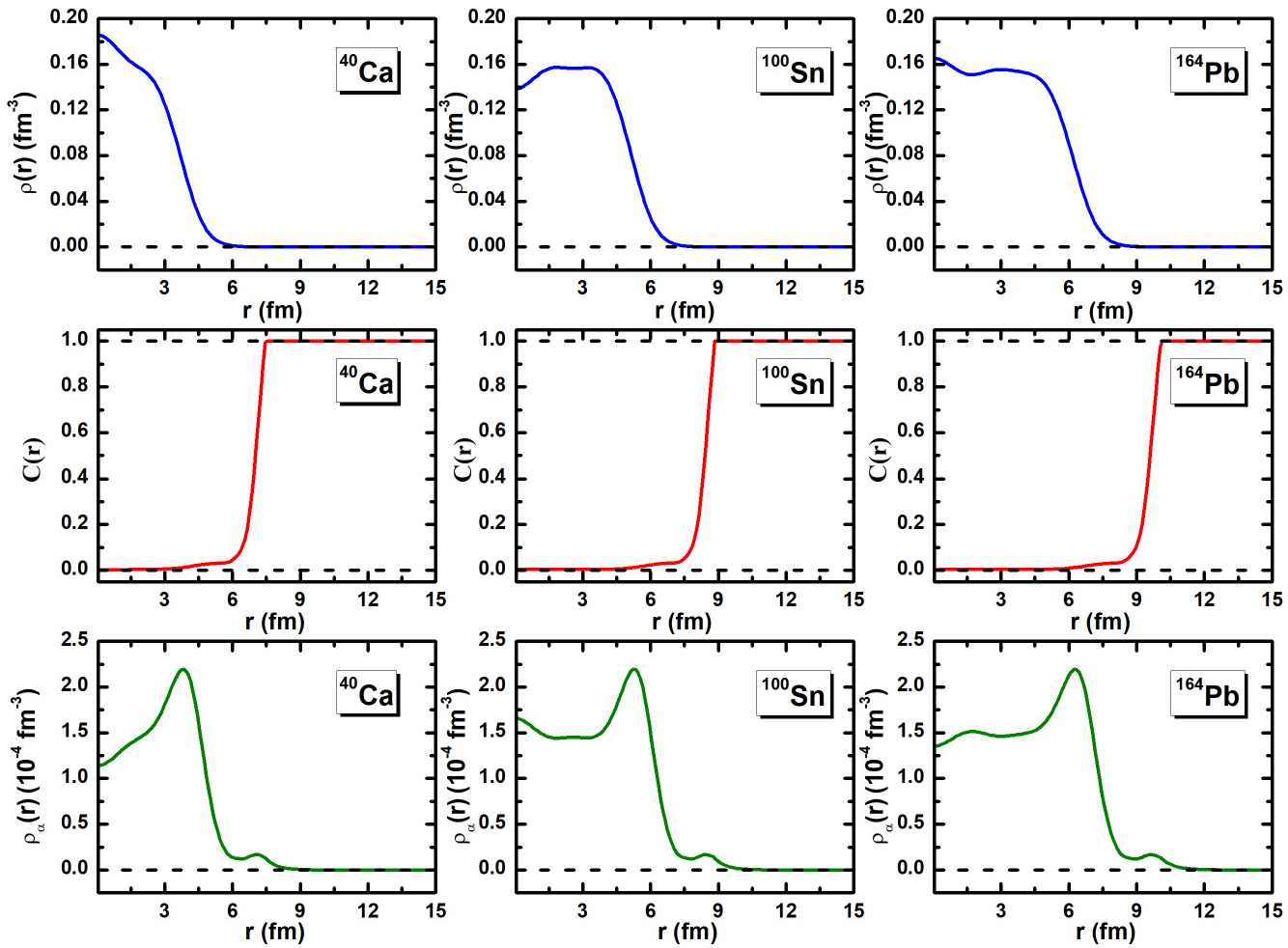}
  \caption{
    Top: Total nucleon density profiles in $ \nuc{Ca}{40}{} $, $ \nuc{Sn}{100}{} $, and $ \nuc{Pb}{164}{} $, respectively.
    Middle: Local quartet condensate fractions $ \mathcal{C} \left( {r} \right) $ as functions of radial coordinate $ r $ in $ \nuc{Ca}{40}{} $, $ \nuc{Sn}{100}{} $, and $ \nuc{Pb}{164}{} $, respectively, within the LDA.
    Bottom: The local quartet condensate density
    $ \rho_{\alpha} \left( {r} \right) = \mathcal{C} \left( {r} \right) \rho \left( {r} \right) / 4 $
    as a function of spatial coordinate $ r $ in $ \nuc{Ca}{40}{} $, $ \nuc{Sn}{100}{} $, and $ \nuc{Pb}{164}{} $, respectively, within the LDA.}
  \label{fig:6}
\end{figure*}
\par
Next, by extending the QBCS theory to the finite nuclei with LDA, we discuss the spatial profile of the quartet condensate fraction and density therein.
The total nucleon density profiles for $ \nuc{Ca}{40}{} $, $ \nuc{Sn}{100}{} $, and $ \nuc{Pb}{164}{} $~\cite{PhysRevResearch.7.013050}
by the Skyrme Hartree-Fock calculation are shown in the top panels of Fig.~\ref{fig:6}.
Based on such total nucleon density profiles, we plot the local quartet condensation fractions $ \mathcal{C} \left( r \right) $
and densities $ \mathcal{C} \left( r \right) \rho \left( r \right) $ as functions of $ r $ through the LDA.
In the middle panels of Fig.~\ref{fig:6},
one can find that $ \mathcal{C} $ rapidly increases at the outer region where $ \rho $ becomes smaller.
In particular, in the outermost region,
$ \mathcal{C} \rightarrow 1 $ can be found.
This indicates that the system prefers the formation of quartet clusters
(i.e., $ \alpha $ particles) in the dilute outer regime of nuclei.
The quartet condensation density shown in the bottom panels of Fig.~\ref{fig:6}
can be obtained as the product of $ \mathcal{C} \left( r \right) $ and $ \rho \left( r \right) $ and thus exhibits a peaked behavior around the surface region of nuclei. 
Such a tendency is consistent with the recent experiments of neutron-rich nuclei~\cite{
  Tanaka2021Science.371.260},
although we do not consider the isospin asymmetry due to the Coulomb interaction for simplicity.
\par
We note that the local maximum of $ \rho_{\alpha} \left( r \right) $ shown in Fig.~\ref{fig:6}
tends to be small compared to the relativistic mean-field (RMF) calculation~\cite{Typel2014Phys.Rev.C89.064321}.
While the $ \alpha $-particle densities obtained by the RMF calculation~\cite{Typel2014Phys.Rev.C89.064321} distribute locally around the surface of nuclei and are strongly quenched in the inner region of the nucleus, our present results exhibit nonzero values of $ \rho_{\alpha} \left( r \right) $ in the inner region. 
The current difference is associated with the fact that the RMF calculation only considers bound $ \alpha $ particles and ignores Cooper quartet correlations above the Mott density of $ \alpha $ particles~\cite{schuck2007quartetting}.
Nevertheless, the total number of quartet condensates $ N_{\alpha} $ shown in Table~\ref{tab:a_w} is in the same order of magnitude with the number of $\alpha$ clusters obtained in the RMF result~\cite{Typel2014Phys.Rev.C89.064321}.
Meanwhile, there would be some potential underestimations of the quartet condensate density in the present framework
  since uncondensed quartets are not taken into account in the current QBCS theory.
  The description would also become relatively worse for lighter nuclei due to stronger inhomogeneity of density, such as a cluster structure, since we perform the extension to finite nuclei with the LDA.
  In addition, the isovector pairing interaction might also need to be considered in order to achieve a more quantitative description.
\par
It should be noted that the quartet density cannot be observed directly in the experiments~\cite{Tanaka2021Science.371.260}.
In this regard, there are still ambiguities about the relationship between the quartet density and experimental observables.
It is necessary to further investigate how the knockout reaction would be affected by the quartet correlations, which could motivate future collaborations between nuclear structure and reaction theories as well as experiments.
\par
In the extension to finite systems, the neutron-rich isotope of each element is regarded as a system consisting of a core (the referenced $ N = Z $ nuclei) and a few extra (residual) neutrons.
As for the neutron-quartet scattering, the polaronic self-energy of extra neutrons can be evaluated as in Eq.~\eqref{eq:e_h}.
The nucleon-quartet correlation contributes to the nuclear equation of state as in Eq.~\eqref{eq27} or \eqref{eq:enucl}, which is proportional to the number of residual neutrons, namely, the asymmetry of nucleons.
Based on that, in the following, we evaluate the Wigner energy of neutron-rich isotopes as a response of $ N = Z $ nuclei when adding residual neutrons.
In detail, we focus on $ \mathrm{Ca} $, $ \mathrm{Sn} $, and $ \mathrm{Pb} $ neutron-rich isotopes, and evaluate the LDA Wigner term originating from the nucleon-quartet correlation as 
\begin{equation}
  E_{\urm{W}}^{\urm{LDA}}
  =
  \int
  W \left( \bm{r} \right)
  \,
  {\Delta \rho_n} \left( \bm{r} \right)
  \, d \bm{r},
\end{equation}
{where $ {\Delta \rho_n \left( \bm{r} \right)} $ is the residual neutron density as introduced in Eq.~\eqref{eq24},
  and here in the finite nuclei it refers to the difference between the density distributions of neutron-rich isotopes and referenced $ N = Z $ core; namely, it is defined as
  \begin{equation}
    \Delta \rho_n \left( \bm{r} \right)
    =\rho_
{\text{total}} \left( \bm{r} \right)
    -
    \rho_{\urm{core}}\left( \bm{r} \right),
  \end{equation}
  where $ \rho_{\text{total}} \left( \bm{r} \right) $ and $ \rho_{\urm{core}} \left( \bm{r} \right) $ are the spatial distributions of the total nucleon density for the neutron-rich isotope and corresponding conjugate $ N = Z $ core, respectively.
}
In addition, one has
\begin{equation}
  W \left( \bm{r} \right)
  =
  \frac{2 \pi a}{M_{\urm{r}}} \rho_{\alpha} \left( \bm{r} \right)
  \equiv
  \frac{\pi a}{2M_{\urm{r}}}
  \mathcal{C} \left( \bm{r} \right)
  \rho_{\urm{core}} \left( \bm{r} \right). 
\end{equation}
By assuming uniform residual neutron density $ \Delta \rho_n \left( \bm{r} \right) $ in the practical calculation, one has
  \begin{equation}
    \Delta \rho_n
    =
    \frac{\left| N - Z \right|}{V},  
  \end{equation} 
  where $ V = 4 \pi R_{\urm{max}}^3 / 3 $ is the volume of the nucleus, and the nuclear size $ R_{\urm{max}} $ is adopted such that $ \rho \left( \left| \bm{r} \right| = R_{\urm{max}} \right) = 10^{-5}\, \mathrm{fm}^{-3} $.
  As a result, we obtain the simplified form
  \begin{align}
    E_{\urm{W}}^{\urm{LDA}}
    & \equiv
      a_{\urm{W}}
      \left| N - Z \right| \notag \\
    & =
      \frac{\pi a}{2M_r}
      \frac{\left| N - Z \right|}{V}
      \int
      \mathcal{C} \left( \bm{r} \right)
      \rho_{\urm{core}} \left( \bm{r} \right)
      \, d \bm{r}.
  \end{align}
We summarize the numerical results of $ a_{\urm{W}} $ in Table~\ref{tab:a_w}.
One can find $ a_{\urm{W}} $ tends to increase for larger nuclei,
indicating that the larger quartet condensation density shown in Fig.~\ref{fig:6} leads to larger $ a_{\urm{W}} $.
\begin{table}[tb]
  \centering
  \caption{
    Wigner term coefficients $a_{\urm{W}}$ and total numbers $N_\alpha$ of quartet condensates of $ \mathrm{Ca} $, $ \mathrm{Sn} $, and $ \mathrm{Pb} $ isotopes obtained by
    QBCS theory and the LDA calculation.
    The empirical values are calculated by $ 47 / A \, \mathrm{MeV} $~\cite{
      Satula1997Phys.Lett.B407.103--109}.}
  \label{tab:a_w}
  \begin{ruledtabular}
    \begin{tabular}{lddd}
      Isotopes & \multicolumn{1}{c}{$ \mathrm{Ca} $} & \multicolumn{1}{c}{$ \mathrm{Sn} $} & \multicolumn{1}{c}{$ \mathrm{Pb} $} \\
      \hline 
      $ a_{\urm{W}}$  ($ \mathrm{MeV} $) & 0.0428 & 0.0523 & 0.0558 \\
      Empirical value ($ \mathrm{MeV} $) & 1.1750 & 0.4700 & 0.2866 \\
      $ N_{\alpha} = \int \rho_{\alpha} \left( \bm{r} \right) \, d \bm{r} $ & 0.1185 & 0.2308 & 0.3601 \\
    \end{tabular}
  \end{ruledtabular}
\end{table}
\par
The empirical values~\cite{
  Satula1997Phys.Lett.B407.103--109,
  KouraPTP,KOURA200047}
for the strength of the Wigner term, $ a_{\urm{W}} = \mathcal{W} / A^{\alpha} $,
extracted from binding energies can be described by the average line with
$ \mathcal{W} = 30 $--$ 50 \, \mathrm{MeV} $ and $ \alpha \simeq 1 $.
By comparing present results with empirical ones, although the mass-number dependence might be different,
it is found that the contribution of the Wigner term from nucleon-quartet cluster scattering is one order of magnitude smaller ($ 4 $--$ 20 \, \% $ depending on $ A $)
of total empirical strength.
Importantly, our results show the same sign as the empirical data.
In this regard, when one tries to fit the empirical value with the phenomenological parameters in the energy density,
the additional quartet contribution shown in this work affects the magnitude of such parameters.
In addition, the total number $ N_{\alpha} $ of quartet condensates is calculated by spatially integrating
$ \rho_{\alpha} \left( \bm{r} \right) $.
It is found that $ 4 N_{\alpha} / A \simeq 0.01 $ holds for all species. 
Both the nucleon-quartet correlation and $ N_{\alpha} $ become relatively larger with increasing mass number.
Our results indicate that the nucleon-quartet correlation may be the partial origin of the Wigner term, whereas its contribution is not sufficient to explain the empirical value.
It should be noted that, in more realistic situations,
the nucleon density involves the isospin asymmetry in the core and moreover the gradient correction beyond the LDA would be worth investigating.
These effects will be further considered in the future work.
{In the present work, while we investigate the nucleon-quartet correlation by focusing on the neutron-rich isotopes, the proton-rich side could be investigated in a similar manner if one ignores the Coulomb interaction.}
\section{Summary and perspectives}
\label{sec:V}
\par
In this paper, we theoretically investigate Cooper quartet correlations in $ N = Z $ nuclei by combining the QBCS theory and the Skyrme Hartree-Fock calculation with the LDA.
Large quartet condensate fractions are found in the low-density regime of infinite symmetric nuclear matter and in the surface region of finite nuclei $ \nuc{Ca}{40}{} $, $ \nuc{Sn}{100}{} $, and $ \nuc{Pb}{164}{} $,
partially manifesting the tendency of $ \alpha $-cluster formation around the surface of medium-mass neutron-rich nuclei~\cite{
  Tanaka2021Science.371.260}.
\par
To see the consequence of quartetting on the nuclear mass formula,
we explore the relation between nucleon-quartet correlations and the Wigner term, whose origin is still elusive.
Based on the fact that the quartet condensate is localized near the surface region with low densities,
we evaluate the effects of the nucleon-cluster repulsive interaction by using the Hartree-Fock approach combined with the nucleon-$ \alpha $ scattering $ T $-matrix.
Eventually, it is found that the contribution to the Wigner term in the mass formula gives about one order of the magnitude of the empirical value.
Our result indicates that nucleon-quartet correlations give a non-negligible contribution to the Wigner term in addition to the neutron-proton pairing previously considered in the literature.
\par
For future perspectives, we have used the non-self-consistent approach for quartet correlations.
It is interesting to see how the nucleon density profiles are further modified by the quartet correlations.
Moreover, the extension to the neutron-rich side and the application to $ \alpha $-knockout reactions is worth investigating for comparison with recent experiments.
\begin{acknowledgments}
  The authors thank Kei Iida, Youngman Kim, and Masaaki Kimura for useful discussions.
  Y.G.~acknowledges support from the Institute for Basic Science (IBS-R031-D1).
  T.N.~acknowledges
  the RIKEN Special Postdoctoral Researcher Program and the JSPS Grant-in-Aid for Scientific Research under Grants No.~JP22K20372, No.~JP23H04526, No.~JP23H01845, No.~JP23K03426, No.~JP23K26538, No.~JP24K17057, No.~JP25H00402, No.~JP25H01558, No.~JP25K01003, and No.~JP25KJ0405.
  H.T.~acknowledges the JSPS Grants-in-Aid for Scientific Research under Grants No.~JP22K13981 and No.~JP23K22429.
  H.L.~acknowledges the JSPS Grant-in-Aid for Scientific Research (S) under Grant No.~JP20H05648.
\end{acknowledgments}
\appendix
\section{Finite-range potential for the $ s $-wave nucleon-$ \alpha $ scattering }
\label{app:a}
\par
In this appendix, we present the construction of the finite-range separable potential for the $ s $-wave nucleon-$ \alpha $ scattering.
The $ T $-matrix reads
\begin{equation}
  \label{eqtm}
  T \left( \bm{k}, \bm{k}'; \omega \right)
  =
  U \left( \bm{k}, \bm{k}' \right)
  +
  \sum_{\bm{p}}
  \frac{U \left( \bm{k}, \bm{p} \right) T \left( \bm{p}, \bm{k}'; \omega \right)}{\omega - p^2 / \left( 2M_{\urm{r}} \right) + i0}.
\end{equation}
\par
For the separable potential $ U \left( \bm{k}, \bm{k}' \right) = U_0 \gamma_k \gamma_{k'} $,
the $ T $-matrix can be written in a separable form as
\begin{equation}
  T \left( \bm{k}, \bm{k}'; \omega \right)
  =
  \gamma_k
  \gamma_{k'}
  t \left( \omega \right).
\end{equation}
Combining with Eq.~\eqref{eqtm}, one obtains
\begin{align}
  t \left( \omega \right)
  & =
    U_0
    +
    U_0
    \sum_{\bm{p}}
    \frac{\gamma_p^2}{\omega - p^2 / \left( 2 M_{\urm{r}} \right) + i0}
    t \left( \omega \right)
    \notag \\
  & \equiv
    U_0
    \left[
    1
    -
    U_0
    \Pi_0 \left( \omega \right)
    \right]^{-1},
\end{align}
where we introduced the two-body propagator
\begin{equation}
  \Pi_0 \left( \omega \right)
  =
  \sum_{\bm{p}}
  \frac{\gamma_{p}^2}{\omega - p^2 / \left( 2 M_{\urm{r}} \right) + i0},
\end{equation}
leading to the onshell value
\begin{equation}
  \Pi_0
  \left( k^2 / \left( 2 M_{\urm{r}} \right) \right)
  =
  -
  \frac{M_{\urm{r}} \Lambda^3}{4 \pi}
  \frac{\left[ 1 + \Lambda \chi \left( \Lambda - 2ik \right) \right] \left( \Lambda + ik \right)^2}{\left( \Lambda^2 + k^2 \right)^2}.
\end{equation}
Consequently, the on-shell $ T $-matrix reads
\begin{align}
  \frac{1}{T \left( \bm{k}, \bm{k}; k^2 / \left( 2 M_{\urm{r}} \right) \right)}
  & =
    \frac{1}{\gamma_k^2}
    \left[
    \frac{1}{U_0}
    -
    \Pi_0 \left( k^2 / \left( 2 M_{\urm{r}} \right) \right)
    \right]
    \notag \\
  & \equiv
    -
    \frac{M_{\urm{r}}}{2 \pi}
    \left(
    -
    \frac{1}{a}
    +
    \frac{1}{2}
    r k^2
    -
    ik
    \right),
\end{align}
\begin{figure}[t]
  \centering
  \includegraphics[width=1.0\linewidth]{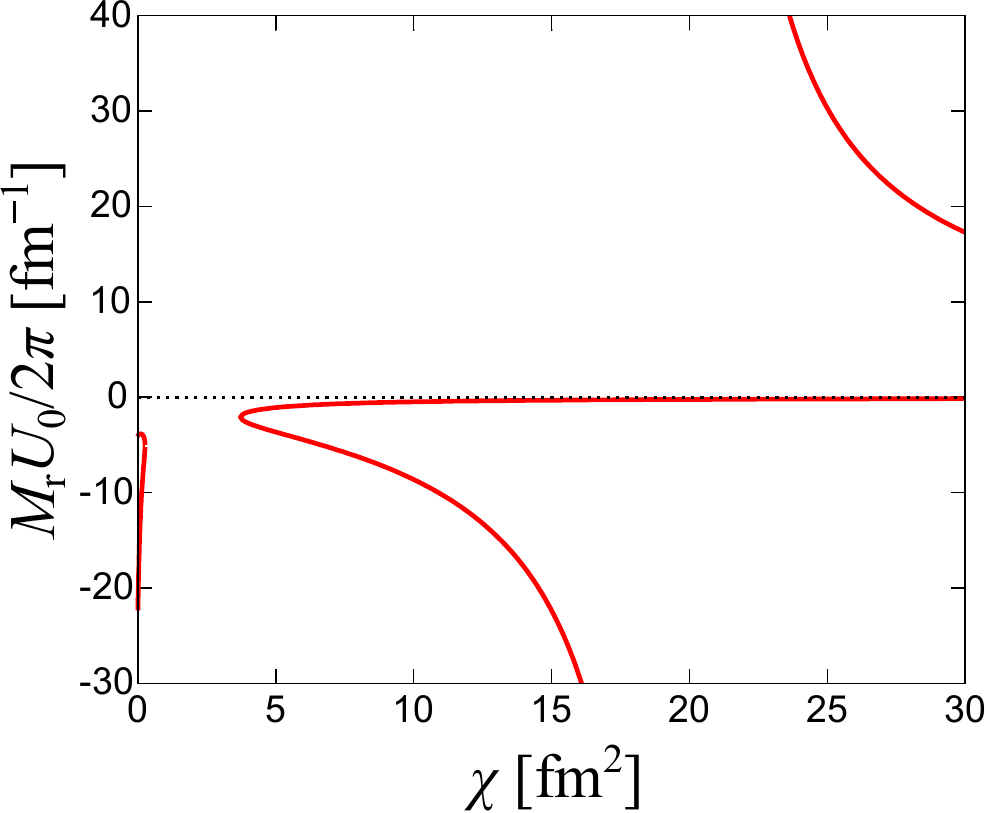}
  \caption{
    The bare coupling constant $ M_{\urm{r}} U_0 / \left( 2 \pi \right) $ as a function of $ \chi $
    in the neutron-$ \alpha $ interaction with the non-monotonic form factor for
    $ a = 2.64 \, \mathrm{fm} $ and $ r = 1.43 \, \mathrm{fm} $.}
  \label{fig:u0}
\end{figure}
Based on the expression above together with
$ 1 / \left( 1 + \chi k^2 \right) = 1 -\chi k^2 + O \left( k^4 \right) $,
the renormalization conditions read
\begin{equation}
  \label{rcu0}
  \frac{1}{U_0}
  +
  \frac{M_{\urm{r}} \left( \Lambda + \Lambda^3 \chi \right)}{4 \pi}
  =
  \frac{M_{\urm{r}}}{2 \pi a},
\end{equation}
and
\begin{equation}
  -
  \frac{M_{\urm{r}} r}{4 \pi}
  =
  \left(
    \frac{2}{\Lambda^2}
    -
    \chi
  \right)
  \frac{M_{\urm{r}}}{2\pi a}
  +
  \frac{M_{\urm{r}} \left( \Lambda^2 \chi - 3 \right)}{4 \pi \Lambda}.
\end{equation}
As a result, one has
\begin{equation}
  r
  =
  -
  \frac{4}{a \Lambda^2}
  +
  \frac{2 \chi}{a}
  +
  \frac{3}{\Lambda}
  -
  \Lambda \chi, 
\end{equation}
From Eq.~\eqref{rcu0}, we can check the sign of $ U_0 $ from
\begin{equation}
  \frac{M_{\urm{r}} U_0}{2 \pi}
  =
  \left[
    \frac{1}{a}
    -
    \frac{\Lambda \left( 1 + \Lambda^2 \chi \right)}{2}
  \right]^{-1}.
\end{equation}
In detail, here we adopt the scattering length $ a = 2.64 \, \mathrm{fm} $
and effective range $ r = 1.43 \, \mathrm{fm} $ for the $ s $-wave scattering amplitude,
which are obtained from the phenomenological $ \alpha $-neutron potential and scattering data in vacuum~\cite{
  Kanada1979Prog.Theor.Phys.61.1327--1341,
  Nakano2020Phys.Rev.C102.055802}.
As a result, the bare coupling constant $ M_{\urm{r}} U_0 / \left( 2 \pi \right) $ is shown as a function of $ \chi $ in the neutron-$ \alpha $ interaction with the nonmonotonic form factor in Fig.~\ref{fig:u0}.
We note that the positive $ U_0 $ can be found only in the region with $ \chi > 20 \, \mathrm{fm}^2 $.
\par 
In the practical calculations, we determined three parameters $ U_0 $, $ \chi $, and $ \Lambda $ by reproducing two scattering parameters $ a $ and $ r $.
In this regard, there is an ambiguity of parameters.
However, due to the low-energy universality, all results are the same in the low-density limit regardless of the specific value for $ \chi $.
%
%

\end{document}